# Implied Open-circuit Voltage Imaging via a Single Bandpass Filter Method – Its First Application in Perovskite Solar Cells

Arman Mahboubi Soufiani[1*], Robert Lee-Chin[1], Paul Fassl[2,3], Md Arafat Mahmud[4,5], Michael E. Pollard[1], Jianghui Zheng[1,4,5], Juergen W. Weber[1], Anita Ho-Baillie[1,4,5], Thorsten Trupke[1], Ziv Hameiri[1]

[1]School of Photovoltaic and Renewable Energy Engineering, University of New South Wales, Sydney 2052, Australia
[2]Light Technology Institute, Karlsruhe Institute of Technology, Engesserstrasse 13, 76131 Karlsruhe, Germany
[3]Institute of Microstructure Technology, Karlsruhe Institute of Technology, Hermann-von Helmholtz-Platz 1, 76344 Eggenstein-Leopoldshafen, Germany
[4]School of Physics, The University of Sydney, Sydney, New South Wales 2006, Australia
[5]The University of Sydney Nano Institute (Sydney Nano), The University of Sydney, Sydney, New South Wales 2006, Australia

E-mail: a.mahboubisoufiani@unsw.edu.au

## Abstract

A direct, camera-based implied open-circuit voltage ($iV_{OC}$) imaging method via the novel use of a single bandpass filter (s-BPF) is developed for large-area photovoltaic solar cells and solar cell precursors. This method images the photoluminescence (PL) emission using a narrow BPF with centre energy in the high-energy tail of the PL emission taking advantage of the close-to-unity absorptivity of typical photovoltaic devices with low variability in this energy range. As a result, the exact value of the sample's absorptivity within the BPF transmission band is not required. The use of a s-BPF enables the adaptation of a fully contactless approach to calibrate the absolute PL photon flux for camera-based spectrally-integrated imaging tools. The method eliminates the need for knowledge of the imaging system spectral response and the use of the emission and excitation spectral shapes. Through an appropriate choice of the BPF centre energy, a range of absorber compositions or a single absorber with different surface morphologies (e.g., planar vs textured) can be imaged, all without the need for additional detection optics. The feasibility of this s-BPF method is first assessed using a high-quality $Cs_{0.05}FA_{0.79}MA_{0.16}Pb(I_{0.83}Br_{0.17})_3$ perovskite neat film. The error in $iV_{OC}$ is determined to be less than 1.5%. The efficacy of the method is then demonstrated on device stacks with two different perovskite compositions commonly used in single-junction and monolithic tandem solar cells.

# Introduction

Fabrication of more efficient and stable single-junction perovskite solar cells (PSC) and monolithic silicon/perovskite tandem solar cells (TSC), both in research laboratories [1]-[6] and industry [7], [8], inevitably requires the development of reliable device characterisation methods. This is especially important when considering the upscaling of these technologies, as large-area solution-processed devices can suffer from increased spatial inhomogeneity of the opto-electronic properties of the constituent layers.

Camera-based luminescence imaging has played a key role in the advancement of silicon solar cell technologies for about two decades [9]-[17] and has more recently found its way to non-silicon photovoltaic (PV) devices [18]–[34]. The open-circuit voltage ($V_{OC}$) and series resistance ($R_s$), which represent the charge-carrier recombination and transport losses throughout the device, respectively, are key solar cell electrical parameters. These two parameters have been spatially quantified using camera-based photo- (PL) and electro-luminescence (EL) imaging techniques [17], [35], [36].

The $V_{OC}$ that is obtained indirectly from measured material or device parameters such as the luminescence signal, is known as the implied-$V_{OC}$ (i$V_{OC}$, i.e., quasi-Fermi level splitting), which represents the highest achievable $V_{OC}$ for a specific sample at a specific processing stage [37]. Depending on the energetic band alignment across the solar cell, the i$V_{OC}$ can deviate from the $V_{OC}$ measured at the device's terminals (herein denoted the "terminal $V_{OC}$") [38], particularly at high illumination intensities [39]. When measured under a wide range of illumination intensities (defined in Suns-equivalent photon flux), the Suns-i$V_{OC}$ curve or Suns-$V_{OC}$ curve when terminal $V_{OC}$ is measured is obtained by plotting the illumination intensity versus the implied or measured voltage, respectively. When compared against the light current-voltage curve of the corresponding device, it allows quantification of $R_s$ [40].

Here, a simple, contactless image calibration approach for *direct* i$V_{OC}$ imaging of PV materials and cells is developed using a single bandpass filter (s-BPF). This removes the need for experimentally quantifying: i) the quantum efficiency of the optical system at the detection side, which can change over time, ii) the full absorptivity spectrum of the sample, and iii) the full spectral shape of the excitation source and sample luminescence. Nevertheless, one needs to ensure that the light transmitted by the chosen narrow BPF resides within the high energy tail of the PL spectrum and, more specifically, where the absorptivity of that light is close to unity and almost constant.

This novel method for i$V_{OC}$ imaging is demonstrated using samples incorporating perovskite compositions commonly used in single-junction and monolithic TSCs. The uncertainties in the i$V_{OC}$ images obtained by the s-BPF method are further evaluated.

## Method Description

Considering the generalised Planck's emission law [37], [41], the absolute PL photon flux per energy interval emitted at different lateral positions on the surface ($\vec{x}$) as a function of photon energy ($\hbar\omega$), is related to i$V_{OC}$ [39], [42] through:

$$\phi_{PL}^{absolute}(\hbar\omega, \vec{x}) = \text{Abs}(\hbar\omega, \vec{x}) \times \phi_{BB}(\hbar\omega) \times \exp\left(\frac{iV_{OC}(\vec{x})}{k_B T}\right) \quad (1.1)$$

assuming a uniform splitting of the quasi-Fermi levels throughout the depth of the absorber. Expanding the blackbody radiation photon flux $\phi_{BB}(\hbar\omega)$:

$$\phi_{PL}^{absolute}(\hbar\omega, \vec{x}) = \text{Abs}(\hbar\omega, \vec{x}) \times C \times (\hbar\omega)^2 \times \exp\left(\frac{iV_{OC}(\vec{x}) - \hbar\omega}{k_B T}\right) \quad (1.2)$$

where Abs($\hbar\omega, \vec{x}$) is the $\vec{x}$-dependent absorptivity of the sample and $\hbar$, $T$, and $k_B$ are the reduced Planck's constant, carrier temperature, and Boltzmann constant, respectively. The second term on the right-hand side of Equation (1.2) [37], [43], [44], assuming an emission into a hemisphere of $2\pi$ sr in front of the sample, is [45], [46]:

$$C = \frac{1}{4 \times \pi^2 \times \hbar^3 \times c^2} \quad (2)$$

where c is the speed of light in vacuum. Note that Equation (1) assumes $\hbar\omega - iV_{OC} \geq 3k_B T$, which is valid for the semiconductors examined in this study. The detected PL intensity, $I_{PL}^{detect}$, depends on the overall spectral response of the measurement setup, including the passive optical components and the detector's sensor. It is linked to the absolute PL flux through an energy (or wavelength)-dependent calibration constant, $k_{cal}(\hbar\omega)$:

$$I_{PL}^{detect}(\hbar\omega, \vec{x}) = k_{cal}(\hbar\omega) \times \phi_{PL}^{absolute}(\hbar\omega, \vec{x}) \quad (3)$$

Strictly speaking, $k_{cal}(\hbar\omega)$ should also depend on lateral position. In imaging systems, for example, one could use the flatfield correction to correct for a spatially dependent calibration constant. Here, the $k_{cal}$ is reasonably spatially invariant (see **Figure S1D**).

The spatial variation of i$V_{OC}$ can then be calculated through a linear fit to the high-energy tail of the left-hand side of the following equation when plotted with respect to photon energy:

$$\ln\left(\frac{I_{PL}^{detect}(\hbar\omega,\vec{x})}{C\times(\hbar\omega)^2\times \text{Abs}(\hbar\omega,\vec{x})\times k_{cal}(\hbar\omega)}\right)=\frac{iV_{OC}(\vec{x})}{k_B T}-\frac{\hbar\omega}{k_B T} \tag{4}$$

where the absorptivity of the sample, in the high-energy end, can be measured reliably using a spectrophotometer or from spectroscopic ellipsometry. $T$ is extracted from the slope of Equation (4) and $iV_{OC}$ from the intercept. This method has been widely applied [46], [47], for instance using high-resolution hyperspectral imaging systems over relatively small areas of less than $500\times 500$ µm$^2$ [42]-[47].

However, spectrally-integrated camera-based imaging is used much more widely for rapid characterisation and inspection of large-area samples, such as >6-inch (larger than 242 cm$^2$) silicon wafers and cells [10], [53]. In this technique, the PL/EL signal is *spectrally integrated* at each image pixel, making the use of Equation (4) appear, at first glance, unfeasible. However, rather than adopting the above commonly practiced method, a simpler, *direct*, and *fully contactless* method is proposed here, that is based on the use of a s-BPF. Combined with Equation (1.1), this eliminates the need to have access to the spectrally resolved luminescence signal at each image pixel for the purpose of $iV_{OC}$ quantification.

The following paragraphs elaborate on adaptation of Equation (1.1) for camera-based, single-shot $iV_{OC}$ image acquisition.

Although the temperature can be easily measured using thermocouples, at the relevant illumination intensities of up to 1-Sun equivalent photon flux, and with relatively short total image acquisition times (≤20 seconds), a significant rise of the sample temperature beyond ambient temperature is not expected [48], [54]. Therefore, a constant $T$ can be accounted for, easing the need for determining the slope of the left-hand side term in Equation (4). The uncertainties caused by variation in $T$ will be discussed below.

Instead of a spectrally resolved absolute PL photon flux we measure a single, spectrally integrated PL image using a narrow bandwidth (13 nm) BPF with sharp cut-on/off edges, the transmission band centred within the high-energy tail of the luminescence spectrum (see **Figure 1**). The need for the filter specification of having a sharp cut-on/off centred at the high energy tail is discussed later in this section.

Only a fraction of the total PL flux emitted by the sample is detected. Light throughput is modulated by the overall optical efficiency (i.e., total transmission and the sensor quantum efficiency) of the detection system and is dominated by the BPF transmission spectrum. The PL flux reaching the monochrome camera sensor is recorded at each pixel as an integration with respect to energy (or, wavelength):

$$\Delta I_{PL}^{detect}(\vec{x}) = \int_{\Delta(\hbar\omega)} I_{PL}^{detect}(\hbar\omega, \vec{x})\, d(\hbar\omega) \tag{5}$$

where $\Delta(\hbar\omega)$ indicates the energy range over which the signal is integrated at sensor pixel. To calibrate for the absolute PL photon flux within the same detection range of the imaging system's optical response, $k_{cal}$ is needed:

$$\Delta\phi_{PL}^{absolute} \approx \Delta I_{PL}^{detect} / k_{cal} \tag{6}$$

Note that the energy dependency of $k_{cal}$ is dropped here since the signal in camera-based imaging systems is integrated. For calibration, a spatially homogeneous emission source with a known absolute spectral photon flux is imaged by the system while the BPF is mounted in front of the camera. In this study, the source is comprised of a halogen lamp [27] that is fibre-coupled into an integrating sphere, for which the absolute spectral photon flux at the integrating sphere output port is known. A detailed description of the calibration procedure is provided in **Note 1** of the Supplementary Information. The output port of the sphere, facing the camera, is imaged at the sample plane. An example of the imaged integrating sphere 3-mm output port is provided in **Figure S1**. The calibration constant of the imaging system is then:

$$k_{cal} \approx \Delta\phi^{detect} / \Delta'\phi^{absolute} \tag{7}$$

where $\Delta'\phi^{absolute}$ is the calculable photon flux within the energy range in the high-transmission part of the BPF (see the Error Analysis section) and $\Delta\phi^{detect}$ is the integrated signal of the calibrated uniform light source detected by the camera. Ideally, a BPF with ultra-narrow bandwidth is required so that in the limit $k_{cal}$ and $\Delta\phi_{PL}^{detect}$ are obtained for a single wavelength. However, this may not be readily accessible, adds constraints on the BPF specifications, and reduces the detected signal, thereby requiring long integration times for acquiring high signal-to-noise ratio images.

It is noteworthy that using Equations (6) and (7) to calculate the absolute PL signal: i) avoids the potential error propagation caused by multiplying the individual optical efficiency of each element in the detection system and ii) means it can be performed as frequently as needed to ensure any changes in the optical response of any of the elements over time are accounted for.

Concerning the absorptivity term in Equation (1.1), a spatially uniform absorptivity can be assumed when the transmission band of the BPF is chosen to be within the spectral range where the absorption is high, close to unity, that is at the high-energy part of the spectrum. This is a reasonable assumption for PV devices for which the absorber thickness (and less likely the composition) does not vary noticeably. We then have:

$$\text{Abs}(\hbar\omega, \vec{x}) \approx \text{Abs}(\hbar\omega) \tag{8}$$

To remove the spectral dependency, the centre energy of the BPF needs to be within the high-energy tail of the PL spectrum. For direct bandgap semiconductors with a sharp and clean absorption edge, such as typical halide perovskites [55], [56], the high-energy tail of the PL spectrum is positioned within the spectral range where the absorptivity of optically optimised layers is large and almost constant. The absorption spectrum can be measured once for each specific sample, allowing the determination of a single constant value that can be used for the BPF detection range. Yet, this one-time measurement is not necessary. For metallised devices and absorber layers thicker than 500 nm, the absorptivity is equal to or larger than 0.85 (examples are given in **Figure S3A** for two widely used perovskite compositions). For non-metallised samples (i.e., device stacks or neat films) with a relatively thick absorber layer (i.e., >500 nm) the absorptivity can be slightly lower, but still fairly constant (see **Figures S3B** and **C** and the indicated absorptivity ranges therein). Note that for thick absorber layers of ~1 µm, the absorptivity of the neat film and non-metallised device stacks approach that of the metallised device stacks at the relevant high-energy part of the PL spectrum (this is further discussed below in the Error Analysis section). Therefore, at the high-energy tail of the luminescence spectrum, considering the generally high external PL quantum yield of halide perovskites, the s-BPF method with a constant absorptivity, $\text{Abs}(\hbar\omega) \approx A$, can be implemented. The further blue-shifted the centre wavelength of the BPF is, the higher and narrower the above-mentioned absorptivity range for neat films and stacks is (see **Figures S3**). As a result, the need for the exact knowledge of the absorptivity i) within the whole PL emission energy range when a long-pass filter is used at the detection side [31], ii) within the high-energy range needed when using Equation (4) [48], and iii) as a measured single-value within the BPF's high-transmission energy range for the generation of $iV_{OC}$ images with the s-BPF method, is eliminated.

In principle, one should shift the transmission band of the BPF as far as possible to high energies, to minimise the risk of getting into the spectral range where the absorptivity starts dropping, however, the further the transmission band is moved to the high-energy range, the lower the PL signal, so eventually it becomes an optimisation of two opposing requirements.

Altogether, through the implementation of the s-BPF method, Equation (1.1) is rearranged into:

$$iV_{OC}(\vec{x}) = k_B T \times \ln\left(\frac{\Delta I_{PL}^{detect}(\vec{x})}{k_{cal} \times A \times \Delta'\phi_{BB}}\right) \quad (9)$$

From Equation (9), $iV_{OC}$ can be readily quantified spatially. $\Delta'\phi_{BB}$ is the spectrally integrated blackbody photon flux within the imaging system's optical efficiency range, which is again dominated by the BPF transmission:

$$\Delta'\phi_{BB} = \int_{\Delta'(\hbar\omega)} \phi_{BB}(\hbar\omega)\, d(\hbar\omega) \quad (10)$$

Although one can in principle calibrate an imaging system by comparison with the measured terminal voltage, using a contactless calibration approach for spectrally-integrated camera-based imaging systems minimises the impact of differences between $iV_{OC}$ and the terminal $V_{OC}$ on the calibration process [38]. Previously observed inconsistencies between the emitted PL signal and the terminal $V_{OC}$ in some perovskite compositions and devices, often linked to ion migration [20], [23], [57], can adversely affect $iV_{OC}$ image calibration that uses contacting. The contactless s-BPF-based image calibration method circumvents such issues.

It is very important to note that the use of the s-BPF would still provide benefits, even if the calibration is done against the measured terminal voltage. This is because one main reason to use such a filter is to make the calibration applicable to all samples which have high absorptivity in the BPF's transmission band, making the calibration more sample independent (e.g., planar versus textured surfaces). For a BPF with centre energy selected at sufficiently high energies of the PL emission tail, the $iV_{OC}$ is unaffected by laterally propagated and scattered PL emission (refer to Ref. [44] for a relevant discussion).

In brief, the introduced s-BPF method does not require a prior knowledge of the optical efficiency of the optical components of the system, nor the spectral information of the sample's absorptivity, nor the use of the spectral shapes of the excitation source and the sample's luminescence spectra. Note that while the exact spectral distribution of absorptivity is not required, one still needs to ensure that it is close to unity and almost constant in the transmission band of the BPF, which is realisable for relevant absorber thicknesses used in efficient PV devices and cell precursors regardless of the sample structure. The s-BPF method does not require knowledge of the absorber bandgap energy nor any secondary calculations such as that of i$V_{OC,rad.}$ as opposed to other reported methods [33]. The only term in Equation (9) that needs to be measured for each sample is $\Delta I_{PL}^{detect}$, while $\Delta '\phi_{BB}$ requires adjustment according to the change in sample temperature.

## Error Analysis

To assess the validity of the s-BPF method, as well as potential sources of error, the i$V_{OC}$ extracted is compared with a reference i$V_{OC}$ obtained from a self-consistent approach (see Supplementary Information **Note 2**). We examine a range of assumptions and simplifications inherent to the method but, for now, neglect the spatial dependency. For this purpose, we use an experimentally measured absolute spectral PL flux (according to Ref. [44]) of a high-quality $Cs_{0.05}FA_{0.79}MA_{0.16}Pb(I_{0.83}Br_{0.17})_3$ (hereafter denoted by B17) neat film at 1-Sun equivalent photon flux (**Figure 1**) with absolute external PL quantum yield (PLQY) of ~25.9%. This Br17 is a well-studied perovskite composition [38], [54]. For the error calculations, a calibrated light-source spectral photon flux measured at the integrating sphere output port is used (**Figure S2B**). Note that in these error calculations no uncertainty associated with the calibrated light-source is considered. Through the multiplication of the corresponding spectral photon flux by the optical efficiency of our custom imaging system (see **Figure S4**), $\Delta I_{PL}^{detect}$ and $\Delta \phi^{detect}$ were calculated. It is important to note that the knowledge of the optical efficiency (i.e., *spectral response*) of the *imaging system* is only used for the error analysis of the s-BPF method and is *not needed* for the actual implementation of the s-BPF method itself.

The two BPFs used here are denoted by BP720/13 (nominal centre wavelength: 720 nm, nominal minimum bandwidth: 13 nm, and effective refractive index of the filter layers $n_{eff}$: 2.04) and BP740/13 (nominal central wavelength: 740 nm, nominal minimum bandwidth: 13 nm, $n_{eff}$: 2.04). **Figure 1** presents the manufacturer-provided transmission spectra. The BPFs' transmission spectra and centre wavelengths were confirmed by in-house measurements (see **Figure S13** and **Note 3** of the Supplementary Information). For the following error analysis, the spectra provided by the manufacturer are used.

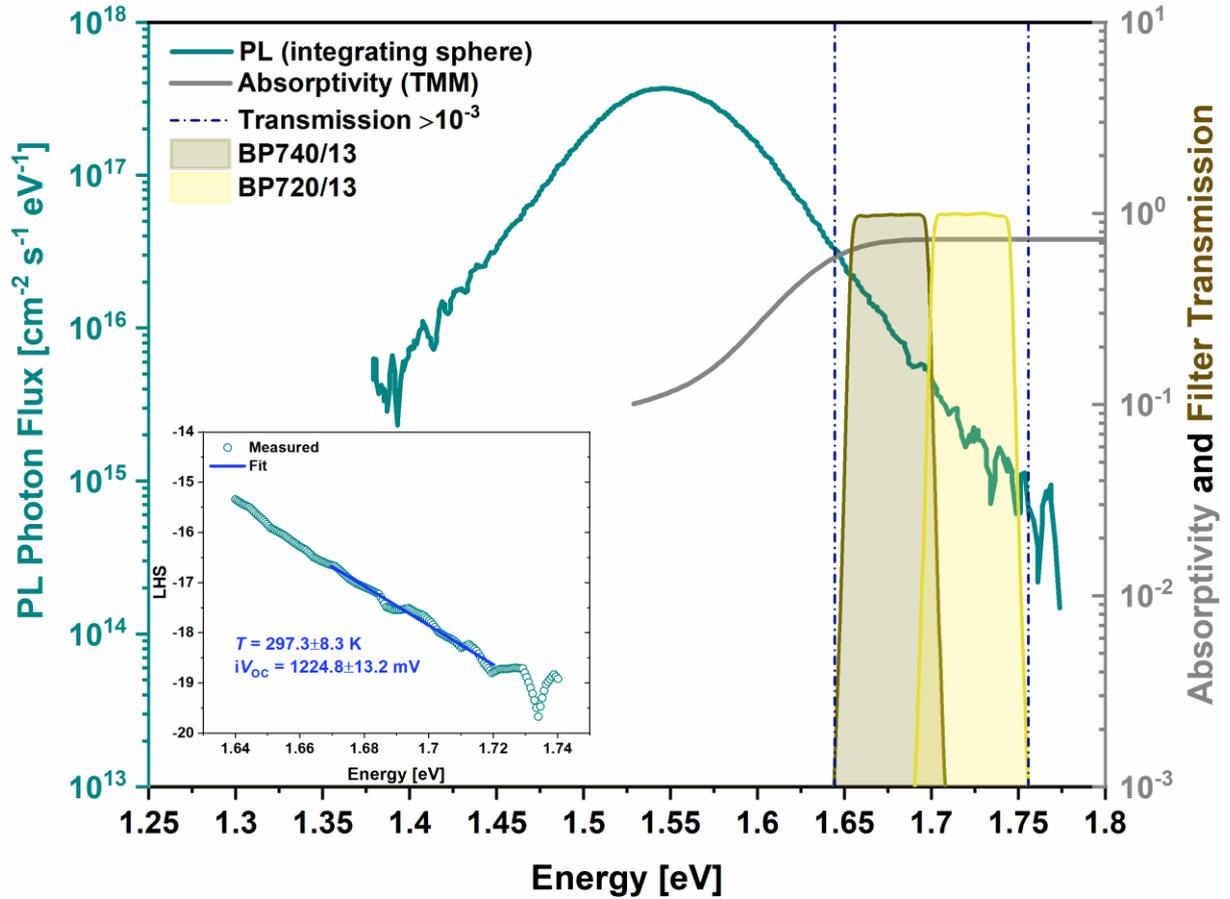

**Figure 1 Sample and filter optical properties and $iV_{OC}$ extraction from absolute spectral PL.** Absolute spectral PL photon flux per energy interval of a $Cs_{0.05}FA_{0.79}MA_{0.16}Pb(I_{0.83}Br_{0.17})_3$ perovskite polycrystalline thin film (thickness: ~471 nm) measured at 1-Sun equivalent illumination intensity. Also presented are the corresponding absorptivity, calculated via the Transfer Matrix Method [58] using the complex refractive index spectra measured by spectroscopy ellipsometry (see Supplementary Information **Figure S8**) as well as the transmission spectra of the two BPFs. The vertical dash-dotted lines mark the energies at $10^{-3}$ transmission for the two BPFs at which the lower and upper limits of the constant absorptivity are set. The insert is an example fit to the experimental spectrum (circular data points) obtained from the high-energy tail of the left-hand side of Equation (4) with two free parameters: $iV_{OC}$ = 1224.8±13.2 mV and $T$ = 297.3±8.3 K. Since this PL spectrum is measured in an integrating sphere, for these calculations, the factor 4 in the physical constant defined in Equation (2) is replaced with a factor 2 so to approximately account for PL emission into the full sphere [47]. The error in these values indicate the 95% confidence interval. Supplementary Information **Note 2** provides further discussion on the impact of the fitting energy range on the extracted $iV_{OC}$ and $T$ and compares the results with the literature [54].

In the following, four potential sources of error are evaluated:

I. <u>Bandpass filter transmission range used for integration:</u>

The stopband transmission of the BPFs used in this study is nominally on the order $10^{-7}$. While for $\Delta\phi^{\text{detect}}$ and $\Delta\phi_{\text{PL}}^{\text{detect}}$ the signal is integrated over the spectral response of the imaging system (at each image pixel), for $\Delta'\phi^{\text{absolute}}$ and $\Delta'\phi_{\text{BB}}$ the integration needs to be performed over the pre-selected energy (or wavelength) range, which falls within the BPF's high-transmission range. The error associated with Equation (9) due to this latter integration range is calculated with respect to the i$V_{\text{OC}}$ of 1224.8 mV, extracted from the fit to the high-energy tail approach at extracted temperature of 297.3 K (see the inset of **Figure 1**). Refer to **Note 2** of the Supplementary Information for a discussion about the extracted i$V_{\text{OC}}$ and *T*.

The wavelength integration interval for $\Delta'\phi^{\text{absolute}}$ and $\Delta'\phi_{\text{BB}}$ is varied within a wide range where the transmission of the BPFs is greater than $10^{-5}$ and less than $10^{-1}$ (see **Figure S11** for the integrated quantities). Absorptivity in Equation (9) is set to a constant value for the example perovskite neat film in **Figure 1**, for which it varies from 0.58 to 0.73. Results for BP740/13 and BP720/13 are provided in **Figures 2A** and **2B**, respectively. The calculated error, at any fixed constant absorptivity within the relevant range (highlighted in **Figure 2**) and at 297.3 K, is less than 0.8% (<10 mV), regardless of the BPF, and is lower for BP720/13 (<0.5%). Note that calculating the error with respect to the i$V_{\text{OC}}$, and the corresponding *T*, obtained from other fitting energy ranges shown in **Figure S6**, results in errors no more than 0.8% and 0.6% for BP740/13 and BP720/13, respectively.

II. <u>Constant absorptivity:</u>

Another simplification implemented to the method is considering a constant absorptivity in the high-energy tail of the PL spectrum. As mentioned above, uncertainties of no more than 0.8% (~10 mV) and 0.6% (~8 mV) are calculated for BP740/13 and BP720/13, respectively (see **Figure 2**). If one, without prior knowledge, just assumes that the absorptivity is large and varies between 0.70 and 0.90, the error would still be <1.1% (<14 mV) and <1.3% (<17 mV) for BP740/13 and BP720/13, respectively.

For standalone films, thicker than the one used for the error calculations in this section, the absorptivity can approach that of a non-metallised and fully metallised device stacks within the high-energy tail of the luminescence spectrum. For instance, using the transfer matrix method (TMM) [58] and the complex refractive index spectra of the Br17 composition (**Figure S8**), absorptivity of 0.86-0.88 is obtained for a 1 µm-thick layer within the wavelength integration range of BP720/13 (see **Figure S15**). Such large absorber thicknesses are commonly used in device fabrication [3], [59]–[61]. Therefore, having an optically thick absorber layer enables the use of a single, large, constant, common *A* value, in Equation (9), for a wide range of structures from neat films to device stacks as well as metallised cells.

Further, the variation in absorptivity is lower, and the constant *A* assumption is therefore more justified, for a suitably selected BPF with a more blue-shifted centre wavelength, even for an intermediate thickness perovskite layer (i.e., 400-500 nm). Hence, the error caused by the constant absorptivity assumption will be even lower. For example, in the case of BP720/13, one should only consider the absorptivity in the 0.70-0.73 range (see **Figure 1A**) giving rise to an uncertainty of less than 0.6% (**Figure 2B**). Therefore, for a sufficiently blue-shifted BPF, a single filter allows characterisation of multiple absorber compositions with different bandgaps using a single common constant *A* (see for instance absorptivity spectra in **Figure S3**). Nevertheless, when using a single BPF to image multiple compositions, from a practical point of view, one should note that acquiring high signal-to-noise ratio images can become challenging under lower photo-excitation intensities for the narrower bandgap material.

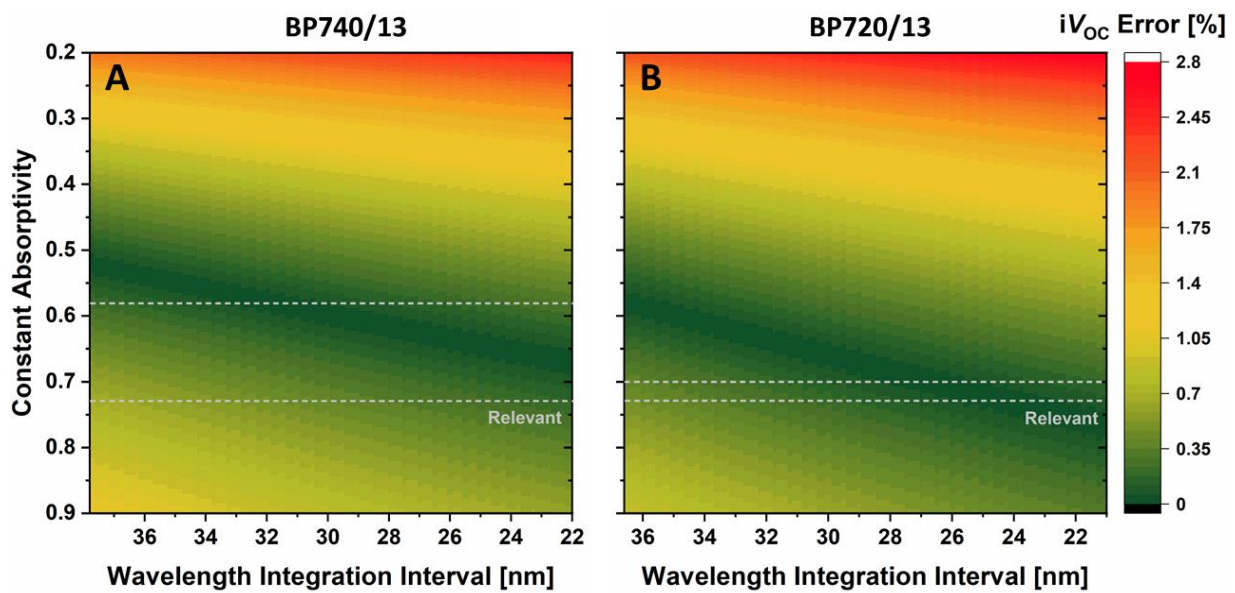

**Figure 2 Method validation and error analysis.** The $iV_{OC}$ relative error maps calculated via the s-BPF method, Equation (9), considering (**A**) BP740/13 and (**B**) BP720/13 filters, with respect to the $iV_{OC}$ of 1224.8±13.2 mV extracted from the fit to the high-energy part of the absolute spectral PL tail at 297.3 K. For this specific perovskite sample (i.e., neat film), absorptivity is varied within the 0.58-0.73 range corresponding to the transmission range >$10^{-3}$ for the two BPFs. The absorptivity bounds, at transmission ≈ $10^{-3}$, for the BPFs are marked with the horizontal dashed lines.

III. Temperature:

In practice, the sample temperature can be controlled using a temperature-controlled stage or can be measured (refer to the PL Imaging Setup of the Characterisation section and **Note 2** in the Supplementary Information for additional discussion). PL imaging with integration times below 20 seconds under illumination intensities of 1-Sun or lower is not expected to significantly raise the sample temperature above the temperature of the measurement environment.

Here, the error in i$V_{OC}$, calculated from Equation (9), due to the deviation of the used temperature from the actual reference temperature, i.e., that obtained from the high-energy fit approach in **Figure 1**, is assessed where the temperature in the s-BPF method is set to a constant value ranging from 294 to 302 K. At a constant absorptivity, the calculated uncertainty remains below 1.1% and 1.3% for BP740/13 and BP720/13, respectively (see **Figure S12**). Note that using different reference i$V_{OC}$ and $T$ values (**Figure S6**), the calculated relative error remains within this limit.

## IV. Angle of incidence dependent wavelength shift of spectral features of the bandpass filter:

Another important consideration for the practicality of this method, when used to image large-area samples, is the spectral blue-shift of the filter optical properties resulting from deviations from the normal incidence of the emitted photons from the sample. This shift increases as the angle of incidence increases i.e., when incident photons on the filter originate from points farther away from the centre of the sample. This spectral shift is reduced as the working distance is increased, since the angles incident on the filter in front of the camera are decreased. Such technical aspects are addressable through the appropriate optical design of the system. The spectral shift can be approximated via [62]:

$$\lambda(\theta) = \lambda_0 \sqrt{1 - \left(\frac{\sin(\theta)}{n_{eff}}\right)^2} \qquad (11)$$

where $\lambda_0$ and $\theta$ are the nominal wavelength of the optical filter and the angle of incidence of the light on the filter, respectively. Since this spectral shift is always towards shorter wavelengths, it pushes the transmission band towards the spectral range of constant absorptivity. This spectral blue-shift can cause an error in the i$V_{OC}$ originating from the calculated $\Delta'\phi^{absolute}$ which is related to the spectral integration range over the BPF transmission (Section I). The choice of a calibrated light source with a constant, or at least slowly changing, spectral photon flux in the vicinity of the filter's high-transmission range can minimise this error.

The error in $iV_{OC}$ caused by this spectral blue-shift for a 5º deviation from the normal incident was calculated. The 5º deviation results in no more than 1 nm blue-shift in the spectral features of both BP720/13 and BP740/13. Note that this is an overestimated angular deviation for our custom-built imaging setup and measurement conditions; the distance between the sample surface and the filter is ~10 cm and the sample area used for the purpose of demonstrating the s-BPF method is about 4 mm × 4 mm. In the relevant absorptivity range, maximum error of 0.8% was obtained for BP740/13 and a lower error of <0.6% for a filter with centre wavelength at higher energy, BP720/13 (see **Figure S12**).

## Results and Discussion

The constituent layers of the inverted device structure used in this study are: glass/ITO/2PACz/perovskite/LiF/C60/BCP. The deposited perovskite layers are either $Cs_{0.05}(FA_{0.85}MA_{0.15})_{0.95}Pb(I_{0.85}Br_{0.15})_3$, hereafter denoted by Br15, or $Cs_{0.05}(FA_{0.77}MA_{0.23})_{0.95}Pb(I_{0.77}Br_{0.23})_3$, hereafter denoted by Br23. The electrical parameters of the associated devices are provided in **Figure S16**. Next, the method is implemented on the corresponding device stacks (non-metallised).

### Implied-$V_{OC}$ Imaging

We first tested the s-BPF method on the Br15-based sample using the two different BPFs, together covering a wide energy range of the high-energy tail of the PL spectrum (see **Figure S3** for the PL and absorptivity spectra). The absorptivity spectra in **Figure S3**, regardless of the perovskite composition, varies between 0.65-0.80 for the non-metallised stacks (absorber layer thickness of 500-600 nm) within the energy range where the transmissivity of both BPFs exceeds $10^{-3}$. We note that even a filter that has a blocking of $10^{-2}$ in the spectral range where the absorptivity drops significantly below unity, would likely still provide accurate $iV_{OC}$ values within the calculated relative error. However, to be conservative, smaller $A$ values (corresponding with larger blocking than $10^{-2}$) were used for the subsequent relative error calculations of the $iV_{OC}$ images (**Table S1**). The raw PL intensity images of the non-metallised device stack measured under 1-Sun equivalent photon flux (with up to ±10% uncertainty; see Characterisation section, PL Imaging Setup, in the Supplementary Information), using BP740/13 and BP720/13 are presented in **Figures 3A** and **3B**, respectively. Some spatial nonuniformities in the PL signal can be seen, rendering it interesting for imaging purposes. The corresponding calculated i$V_{OC}$ images based on the procedure described above are provided in **Figures 3C** and **3D**. Importantly, considering the relative error of less than 1% caused by the measurement uncertainties (see **Figure S19**), the area-averaged i$V_{OC}$ values obtained from the two BPFs are, within uncertainty, in good agreement with each other (1091.9 and 1094.8 mV by BP740/13 and BP720/13, respectively). These values are reported at 298 K (refer to the Characterisation section, PL Imaging Setup, in the Supplementary Information for further discussion on the sample temperature). The relative error images (less than 10 mV) associated with the measurement uncertainties in each parameter of Equation (9) for the two BPFs used for Br15 imaging are provided in **Figures S19A** and **B**.

The i$V_{OC}$ values obtained from the s-BPF method are compared with those calculated from the measured absolute PLQY on the same samples. The latter approach, requiring the calculation of the radiative limit of i$V_{OC}$ for any specific sample, resulted in i$V_{OC}$ values about 40-50 mV higher than those obtained by the s-BPF method. For further discussion about this discrepancy and detailed explanation of how the PLQY-based calculations were performed, refer to **Note 5** of the Supplementary Information.

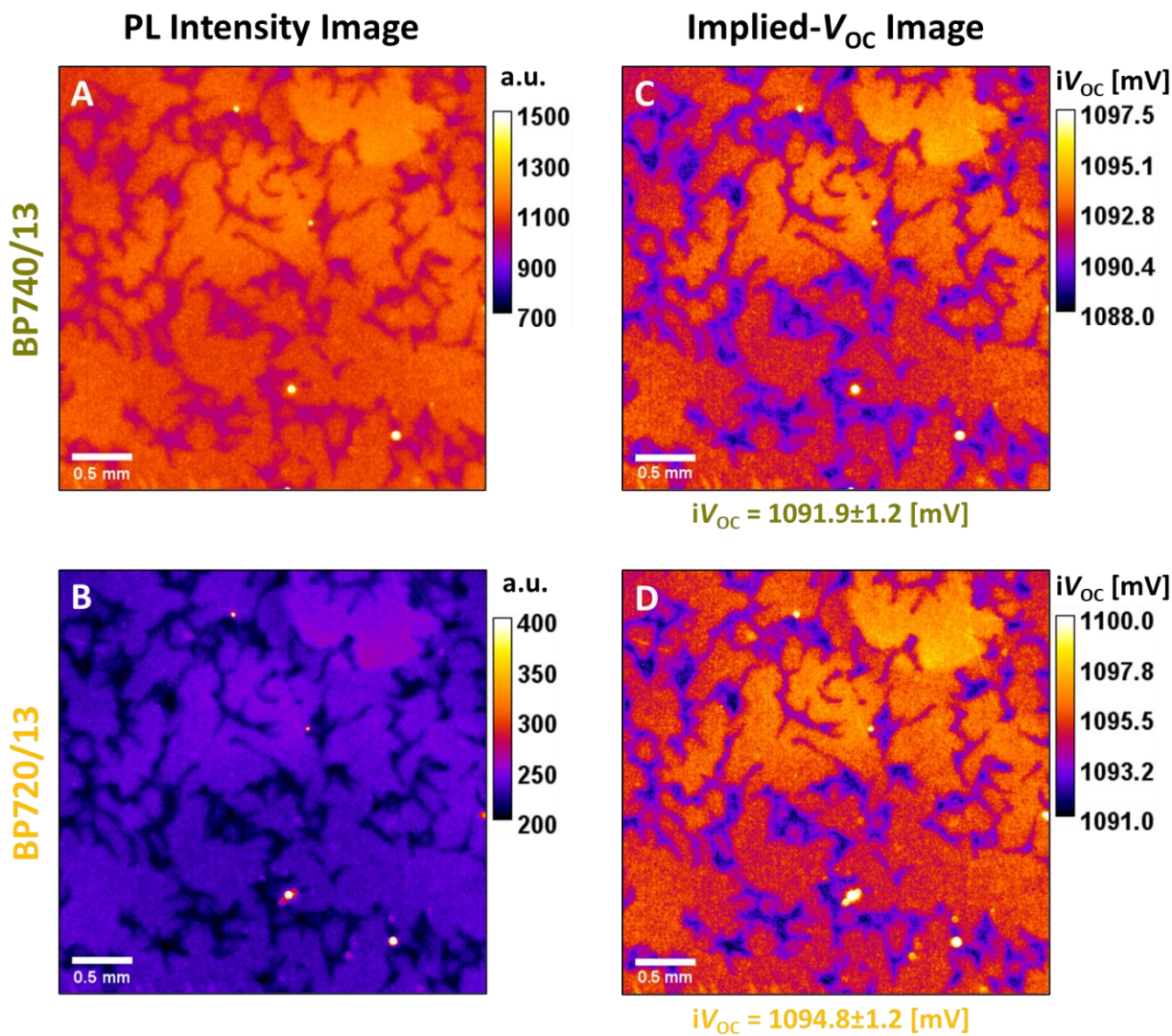

**Figure 3 PL intensity and i$V_{OC}$ images of a Cs$_{0.05}$(FA$_{0.85}$MA$_{0.15}$)$_{0.95}$Pb(I$_{0.85}$Br$_{0.15}$)$_3$ based non-metallised device stack.** Panels **(A)** and **(B)** are the PL intensity images (in arbitrary units) of the device stack acquired using the BP740/13 and BP720/13 filters, respectively. The presented images are acquired with 1.5 s exposure time and 10-times averaging. See Supplementary Information **Note 4** regarding the impact of exposure time on the PL signal. Panels **(C)** and **(D)** are the respective i$V_{OC}$ images calculated through the s-BPF method using the absorptivity of 0.8 (**Figure S3B**) and $T$ = 298 K in Equation (9). The terminal $V_{OC}$ of the corresponding metallised devices is 1036.0±15.2 mV. The standard deviations provided under the i$V_{OC}$ images represent the non-uniformity in the corresponding images. Refer to **Figure S19** for the assessment of the impact of measurement uncertainties on the quantified i$V_{OC}$ images. The histograms associated with i$V_{OC}$ distribution in panels **(B)** and **(C)** are shown in **Figure S21**.

The BP720/13 filter was then used to test the method on the stack of a wider bandgap perovskite absorber, Br23. The raw PL image measured under 1-Sun equivalent photon flux (up to ±10% uncertainty) is presented in **Figure 4A**. The corresponding calculated i$V_{OC}$ image is provided in **Figure 4B**, with an area-average i$V_{OC}$ of 1123.5 mV. The relative error image associated with the measurement uncertainties in each parameter of Equation (9) for the BP720/13 used for Br23 imaging is provided in **Figure S19C**. This error is within ±1%.

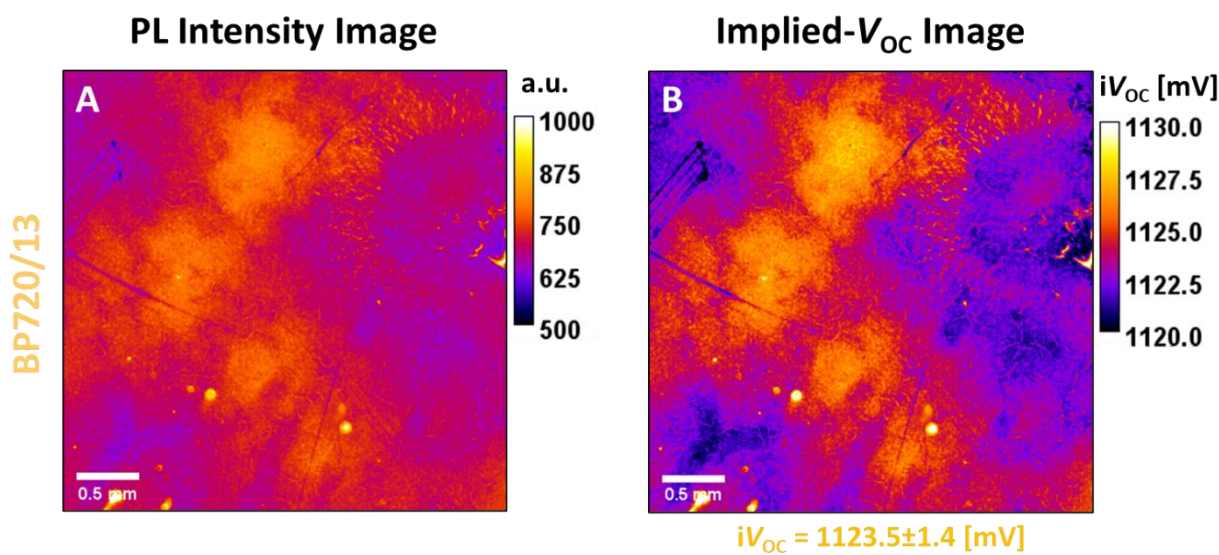

**Figure 4 PL intensity and i$V_{OC}$ images of a Cs$_{0.05}$(FA$_{0.77}$MA$_{0.23}$)$_{0.95}$Pb(I$_{0.77}$Br$_{0.23}$)$_3$ based non-metallised device stack.** Panel **(A)** represents the PL intensity image (in arbitrary units) of the device stack acquired using the BP720/13 filter. The presented image is acquired with 1.5 s exposure time and 10-times averaging. See Supplementary Information **Note 4** regarding the impact of exposure time. Panel **(B)** is the corresponding i$V_{OC}$ image calculated through the s-BPF method using the $A$ = 0.8 (**Figure S3B**) and $T$ of 298 K in Equation (9). The terminal $V_{OC}$ of the corresponding metallised devices is 1090.4±8.8 mV. Refer to **Figure S19** for the assessment of the impact of measurement uncertainties on the quantified i$V_{OC}$ image. The histogram associated with i$V_{OC}$ distribution in panel **(B)** is shown in **Figure S21**.

These results indicate that using just a single bandpass filter, here BP720/13, one can readily spatially quantify the i$V_{OC}$ of perovskite compositions with different absorption thresholds using imaging techniques which, for instance, are deployed in the fabrication of high-efficiency single-junction and monolithic silicon/perovskite TSCs. As such, the potential complication associated with using multiple detection filters in a PL imaging system is eliminated when characterising absorbers with different compositions, or a single composition but with different surface morphologies, allowing high-throughput inspection at an industrial scale.

## Conclusions and Outlook

A simple, fully contactless method that only requires the selection of a single bandpass filter was developed for *direct* i$V_{OC}$ image quantification of photovoltaic devices and solar cell precursors with no need for a priori knowledge of the system's optical response, spectral information of the sample absorptivity, or the radiative limit of the sample's $iV_{OC}$. The relative errors in the resulting absolute $iV_{OC}$ data, associated with the assumptions and simplifications of the proposed method, are within ±1.5%. It is noteworthy, that $iV_{OC}$ variations across the sample, are not associated with any significant error, and accurately reflect lateral variations in sample properties, independently of the calibration procedure that is used to convert PL intensities into implied voltages. This is an inherent benefit of PL imaging-based $iV_{OC}$ quantification.

It was demonstrated that an appropriate choice of the bandpass filter, for which the centre energy of the transmission band is located within a sufficiently high energy part of the PL tail, allows accurate $iV_{OC}$ image quantification for perovskites with various compositions (i.e., different bandgaps) using only a single bandpass filter and a *common*, *constant*, *large* absorptivity value.

Importantly, the use of the s-BPF would still provide benefits, even if the calibration is done against the measured terminal voltage. This is because a main reason to use such a filter is to make the calibration applicable to all samples which have high absorptivity in the BPF's transmission band, making the calibration more sample independent (i.e., samples with different surface morphologies for instance planar versus textured surfaces). This is because for a BPF with centre energy selected at sufficiently high energies of the PL emission tail, the $iV_{OC}$ is unaffected by laterally propagated and scattered PL emission.

The method can readily be extended to contacted mode measurements, such as EL imaging, and may also be exploited for other direct bandgap semiconductors with large Urbach energies or indirect bandgap semiconductors, which may need bandpass filters with centre energy pushed to high-enough energies to ensure large and almost constant absorptivity.

Through this method, it is feasible to directly image $iV_{OC}$ of the sub-cells in monolithic TSCs. When implemented on the silicon bottom-cell, samples with a textured surface are preferred for which the high absorptivity condition of the low-energy photons transmitting through the top-cell in the filter's transmission band is more readily met without needing long integration times to acquire images with a high signal-to-noise ratio. If sufficiently blue-shifted, so the close-to-unity absorptivity condition can be realised for planar silicon samples with no light trapping mechanism, a s-BPF can be used to image $iV_{OC}$ of both planar and textured silicon samples [63] uninfluenced by, for instance, light scattering.

Finally, we note that further adaptation of this method for imaging industrial-size solar cells (>6-inch cells) will require tailoring the optical design of the imaging system, mainly to minimise the errors in the $iV_{OC}$ toward the edges of the devices that are associated with the blue-shift in the bandpass filter's transmission for large angles of incidence. A simple way to achieve that is to increase the working distance between the sample and the camera and choosing an imaging lens with sufficiently long focal distance.

## Authors Contribution

A.M.S. conceived the idea, which was further developed in discussion with T.T. and Z.H. R.L.C. assisted A.M.S. with data analysis and proof-reading the first draft of the manuscript. The samples were fabricated by M.A.M. and J.Z. The absolute spectral PL was measured by P.F. The technicality of the optics of the system was developed by A.M.S. and J.W. and the calibration procedure by A.M.S. and M.P. The first draft of the manuscript was prepared by A.M.S. All authors discussed the results and contributed to the preparation of the final manuscript.

## Acknowledgement

A.M.S. would like to thank Soma Zandi for measuring EQE on Br15- and Br23-based devices, Dr. Yajie Jessica Jiang for the assistance with fitting the spectroscopy ellipsometry data, Dr. Oliver Kunz and Dr. Saman Jafari for constructive discussions in and out of the lab, and Jihoo Lim for contributing to sample fabrication during the initial stages of the study. This work has been supported by the Australian Government through the Australian Renewable Energy Agency (ARENA, Grant RND 2017/001) and the Australian Centre of Advanced Photovoltaics (ACAP). The views expressed herein are not necessarily the views of the Australian Government, and the Australian Government does not accept responsibility for any information or advice contained herein. A.M.S. acknowledges the funding support from ACAP (RG193402-I).